\documentclass[10pt,letterpaper]{article}
\usepackage[margin=2.5cm]{geometry}
\usepackage[affil-it]{authblk} 
\usepackage[utf8]{inputenc} 
\usepackage{float}
\usepackage{relsize}
\usepackage{amsmath}
\usepackage{amssymb,breqn,physics}  
\usepackage{graphicx}
\usepackage{xcolor}
\usepackage{caption}

\usepackage{fancyhdr}
\usepackage{abstract}

\numberwithin{equation}{section}

\usepackage[hidelinks]{hyperref}
\hypersetup{
    citecolor=black, 
    colorlinks = true,
    urlcolor={black},
    linkcolor=.  
    }

\makeatother
\definecolor{hyperlink}{rgb}{0,0,95}  
\usepackage[
    backend=biber,
    style=numeric-comp,
    sorting=none 
]{biblatex}
\title{\bfseries{ Constant-Roll Inflation in modified $f(R,\phi)$ gravity model using Palatini Formalism}}
\author{Sukanta Panda\footnote{Electronic address: \href{mailto:xyz1@gmail.com}{ sukanta@iiserb.ac.in} }, Arun  Rana\footnote{Electronic address: \href{mailto:xyz2@iiserb.ac.in}{ arunrana@iiserb.ac.in} }, Rahul Thakur\footnote{Electronic address: \href{mailto:xyz3@iiserb.ac.in}{ rahul19@iiserb.ac.in}}}
\date{\today}
\affil{Department of Physics, Indian Institute of Science Education and Research Bhopal, Bhopal, Madhya Pradesh, India}

\renewcommand{\thesection}{\Roman{section}.} 
\renewcommand{\thesubsection}{\Alph{subsection}.} 
\usepackage[explicit]{titlesec}
\titleformat{\section}
{\normalfont\normalsize\bfseries}{\thesection}{12pt}{\filcenter
\MakeUppercase{#1}} 
\titleformat{\subsection}
{\normalfont\normalsize\bfseries}{\thesubsection}{12pt}{\filcenter{#1}} 

\begin{document}
\captionsetup[figure]{labelfont={up},labelformat={default},labelsep=colon, name={FIG.}} 

\maketitle 
\begin{abstract}
 
 In this work, we study a constant-roll inflationary model in the Palatini formalism using modified gravity. Here our action consists  a non-minimal coupling of a scalar field $\phi$ with Ricci scalar $R$ in a general form of $f(R,\phi)$. Using Palatini approach, we write its equivalent scalar-tensor form in the Einstein frame and then apply the constant-roll condition in the equation of motion for the inflaton field. Later the tensor-to-scalar ratio and the spectral index are calculated using the slow-roll parameters and the results obtained are matched with the Planck 2018 data. We found that the results agree nicely with the observations within the parameter regime under consideration.
 

\end{abstract}
\vspace{1em} 

\section{Introduction}

Inflation \cite{Guth:1980zm,STAROBINSKY198099,Linde:1981mu,LINDE1983177,Lyth:1998xn,Riotto:2002yw,Baumann:2009ds} is the period of exponential expansion of the universe in the first few instants when it was born. It was first introduced in the 1980s to resolve the problems of the standard Big Bang cosmology like the horizon problem and the flatness or the fine-tuning problem. Not only did it help in resolving these problems, it also provided a mechanism for generating the primordial fluctuations which might have resulted into the large scale structure of the universe that we see today \cite{Mukhanov:1981xt,Mukhanov:1982nu, Starobinsky:1982ee,PhysRevLett.49.1110,Brandenberger:1986pj,PhysRevD.28.679}. One other prediction was the scale-invariance of the primordial power spectrum of cosmological fluctuations which has been confirmed in the observation of the Cosmic Microwave Background (CMB) anisotropies \cite{WMAP:2003ivt,WMAP:2006bqn,Planck:2018vyg}. Though inflation provides an excellent description of the early universe, different alternative models have been proposed based on bouncing cosmology \cite{Novello:2008ra,Nojiri:2017ncd} which attempt to provide a viable non-singular description of the universe \cite{Cai_2013}. One such model is of matter-dominated collapsing universe \cite{Wands_1999,Finelli:2001sr} which is succeeded by the non-singular bounce and can act as an alternative to inflation while also explaining the scale-invariance of the primordial power spectrum \cite{Brandenberger:2009jq,Brandenberger:2009jq,Brandenberger:2012zb,Li:2020pww,Modan:2022fzg}. Though no one can exclude the possibility of a cyclic evolution of the universe, it is much easier to work and to understand the theory of inflation starting from Big Bang imagining a point-sized universe with an infinite density and temperature with all the fundamental interactions unified by a yet unknown framework.

One of the popular models of the inflation is the Starobinsky model \cite{Starobinsky:1980te} which has an additional $R^2$ term in the Einstein-Hilbert action in a pure gravity scenario and is in agreement with the observations in regards to its inflationary spectrum. For other modified gravity models, one can look into 
 \cite{DeFelice:2010aj,Sotiriou:2008rp,Capozziello:2011et,Clifton:2011jh,Odintsov:2016vzz}. In such models, the scalar degree of freedom (inflaton) arises as the effective scalar degree of freedom induced from the gravity and one can recast the action $\int\mathrm{d}^4 x \sqrt{-g}f(R)$ as a scalar-tensor theory with a non-minimal coupling of the aforementioned effective scalar degree of freedom with the Ricci Scalar. But this entire analysis works best in the metric formulation of General Relativity (GR). There is another formulation in GR which is widely popular and is known as Palatini formalism which treats the space-time connections as an independent variable \cite{Sotiriou:2006hs,Sotiriou:2006qn, Borunda:2008kf,Olmo:2011uz} while this not the case for the metric approach. Even though both approaches lead to the same equations for the Einstein-Hilbert action, for modified gravity models and the models where the fields are non-minimally coupled to gravity, this statement is not valid. Here, both these formalisms present different physical scenarios. The Starobinsky model, falling under the general category of the $f(R)$ gravity theories, when studied in the Palatini formalism, leads to no inflationary solutions due to the absence of an additional propagating degrees of freedom which is there if we study the system via the metric approach. Hence to study the inflation in Palatini formalism, a scalar field needs to be added to the action and the first such case was studied in \cite{Bauer:2008zj} where the non-minimal coupling of the scalar field was considered. Since then, many studies have been done considering different inflationary potentials, reheating, post-inflationary phases, preheating and dark matter using $R^2$ term in the action \cite{Tamanini:2010uq,Rubio:2019ypq,Giovannini:2019mgk,Bostan:2019wsd,Gialamas:2019nly,Shaposhnikov:2020fdv,Bauer:2010jg,Kozak:2018vlp,Jinno:2018jei,Lloyd-Stubbs:2020pvx,,Sa:2020qfd,Antoniadis:2020dfq,Ghilencea:2020piz,c,Rigouzzo:2022yan,Dioguardi:2021fmr,Cheong:2021kyc,Lykkas:2021vax,Gialamas:2020vto,Rasanen:2018ihz,Racioppi:2018zoy,Bombacigno:2018tyw,Antoniadis:2018ywb,Antoniadis:2018yfq,Kannike:2018zwn,Almeida:2018oid,Tenkanen:2019jiq,Enqvist:2011qm,Borowiec:2011wd,Stachowski:2016zio,Fu:2017iqg,Rasanen:2017ivk,Racioppi:2017spw,Markkanen:2017tun,Jarv:2017azx,}. For more detailed review on Palatini formalism, one can look at \cite{Tenkanen:2020dge}.

Usually, a scalar field, called as `inflaton' drives the exponential expansion of the universe during the inflationary period with the help of an approximately flat potential. This inflaton field slowly rolls down the potential till the minimum is reached thus ending the inflation. This scenario is called slow-roll inflation and one can look at \cite{Linde:1981mu,Noh:2001ia,Baumann:2009ds,Dioguardi:2021fmr,Riotto:2002yw} for more details. There are other inflationary solutions as well where this slow-roll condition can be ignored. The `ultra slow roll inflation' \cite{Tsamis:2003px,Dimopoulos:2017ged,Pattison:2018bct} is one such scenario in which the curvature perturbation are not frozen at the super Hubble scales thus leading to non-Gaussianities and a more generalized version of that is known as `constant-roll inflation' \cite{Yi:2017mxs,Anguelova:2017djf,Cicciarella:2017nls,Guerrero:2020lng} which consider the rate of acceleration and velocity of the inflaton field as constant. Even though the constant roll models seem to be close to the slow roll ones but these models help us in calculating the slow roll parameters  more precisely. Also, another important prediction of the constant roll inflation is the presence of larger non-Gaussianities \cite{Namjoo:2012aa} in the bispectrum than the ones predicted by the ultra-slow roll case. 

In this paper, we are exploring the constant roll inflation via Palatini formalism when the Einstein-Hilbert action has an additional $R^2$ term non-minimally coupled with a scalar field where $R$ is Ricci scalar. We begin by first finding the scalar field equation for action given in  \cite{Das:2020kff} in Sec.\ref{sec:Model}. Then we proceed to find slow roll parameters in Sec.\ref{sec:cri} using the constant-roll condition. In Sec.\ref{sec:cri}\ref{results}, we plot our results obtained for the spectral index $(n_s)$ and tensor-to-scalar ratio $(r)$ and do a comparison with the Planck data \cite{Planck:2018vyg}. Finally, we conclude with Sec.\ref{sec:conclusion} and summarize our findings with the future direction.


\section{$R^2$ coupled gravity in the Palatini formalism}\label{sec:Model}
We start with the following action\cite{Das:2020kff}, 

\begin{eqnarray}
S &=& \int\mathrm{d}^4 x \sqrt{-g} \left[  \frac{1}{2}f(R,\phi) - \frac{1}{2}g^{\mu \nu}  \partial_{\mu} \phi \partial_{\nu} \phi - V(\phi) \right] \label{2.1},
\end{eqnarray}
where $g_{a b}$ and $g$ are the  metric and its determinant respectively,  $f(R,\phi)=G(\phi)(R+\alpha R^2)$\cite{Rador:2007wq} and $\alpha$ is constant. Here $R$ is the Ricci scalar, which is defined as $R=g^{a b} R^{c}_{a b c}(\Gamma, \partial \Gamma)$ in the Palatini formalism and we assumed the Planck mass to be 1. This action (\ref{2.1}) can be equivalently expressed in terms of an auxiliary field $\chi$ as, 
\begin{eqnarray}\label{2.2}
S &=& \int\mathrm{d}^4 x \sqrt{-g} \left[  \frac{1}{2} f(\chi,\phi) +  \frac{1}{2}\ f'(\chi,\phi) (R - \chi) \ - \frac{1}{2} g^{\mu \nu} \partial_\mu \phi \partial_\nu \phi - V(\phi) \right] \ ,
\end{eqnarray} 
where $f'(\chi,\phi)=\frac{\partial f(\chi,\phi)}{\partial \chi}$, satisfying $\chi=R$ on-shell. Thus  we can rewrite eq.(\ref{2.2}) as, 
\begin{eqnarray}\label{2.3}
S &=& \int\mathrm{d}^4 x \sqrt{-g} \left[  \frac{1}{2} f'(\chi,\phi) R - W(\chi,\phi)- \frac{1}{2} g^{\mu \nu} \partial_\mu \phi \partial_\nu \phi - V(\phi) \right] \ ,
\end{eqnarray}
where $W(\chi,\phi)= \frac{1}{2}\chi f'(\chi,\phi) - \frac{1}{2} f(\chi,\phi)$. Now using the  conformal transformation  
\begin{eqnarray}
g_{\mu \nu} \rightarrow f'(\chi,\phi)  g_{\mu \nu} \ ,
\end{eqnarray}
action (\ref{2.3}) can be written  in Einstein frame as
\begin{eqnarray}\label{action4}
S &=& \int\mathrm{d}^4 x \sqrt{-g} \left[  \frac{1}{2} R -\frac{1}{2}\frac{ \partial_\mu \phi \partial^\mu \phi}{f'(\chi,\phi)}- \frac{W(\chi,\phi)+V(\phi)}{f'(\chi,\phi)^2} \right] \  .
\end{eqnarray}
Now defining a new potential $\hat V(\chi,\phi)$ as
\begin{eqnarray}\label{V}
\hat V(\chi,\phi)\equiv \frac{W(\chi,\phi)+V(\phi)}{f'(\chi,\phi)^2},
\end{eqnarray}
and by using $f(R,\phi) = G(\phi)(R+\alpha R^2)$, the above equation can be written as 
\begin{eqnarray}\label{V1}
\hat V(\chi,\phi)= \frac{1}{f'(\chi,\phi)^2} \left[\frac{1}{8 \alpha G(\phi)}[f'(\chi,\phi)-G(\phi)]^2 +V(\phi)\right] .
\end{eqnarray}
Now varying the action (\ref{action4}) with respect $\chi$, we get
\begin{eqnarray}\label{V2}
f'(\chi,\phi_{})= \frac{8\alpha V(\phi) + G(\phi)}{1- 2\alpha \partial_\mu \phi \partial^\mu \phi} .
\end{eqnarray}
Inserting equation (\ref{V2}) into equation (\ref{V1}), $\chi$ could be eliminated. And then substituting equation (\ref{V1}) into equation (\ref{action4}) and simplifying, we obtain   
\begin{equation}
    S=\int\,d^4x \sqrt{-g} \left[\frac{R}{2}-\frac{1}{2}\frac{1}{(8\alpha V+G)} \partial^{\mu}{\phi}\partial_{\mu}{\phi} +\frac{\alpha}{2}\frac{1}{(8\alpha V+G)} (\partial^{\mu}{\phi}\partial_{\mu}{\phi})^2 -\frac{V}{G(8\alpha V+G)}\right],
\end{equation}
with the effective Lagrangian \cite{Antoniadis:2020dfq}
\begin{equation}
  \mathcal{L}(\phi ,X)\equiv A(\phi)X+B(\phi)X^2-U(\phi), 
\end{equation}
where \begin{equation}\label{2.11}
  X \equiv \frac{(\nabla{\phi})^2}{2} \quad and \quad
\begin{cases}
A(\phi) \equiv \frac{1}{G(\phi)+8 \alpha  V(\phi)} = \frac{1}{G(\phi)}\left(1+8\alpha \frac{V(\phi)}{G(\phi)}\right)^{-1}, \\
   B(\phi) \equiv \frac{2 \alpha }{G(\phi)+8 \alpha  V(\phi)}=\frac{2\alpha}{G(\phi)}\left(1+8\alpha \frac{V(\phi)}{G(\phi)}\right)^{-1}, \\
    U(\phi) \equiv \frac{V(\phi)}{G(\phi) (G(\phi)+8 \alpha  V(\phi))}=\frac{V(\phi)}{G(\phi)^2}\left(1+8\alpha \frac{V(\phi)}{G(\phi)}\right)^{-1}
    \end{cases}.
\end{equation}\\
Here $\mathcal{L}$ contains up to quartic kinetic terms with field-depended coefficients, which belong to k-inflation models and $U(\phi)$ is the effective potential. 
By varying the action with respect to metric $g_{\mu \nu}$, the energy momentum tensor of the source field $\phi$ is written as,
 \begin{equation}
   T_{\mu \nu}=-\frac{2}{\sqrt{-g}}\frac{\delta S}{\delta g^{\mu \nu}}=-(A+2BX)\nabla_{\mu}\phi \nabla_{\nu} \phi + g_{\mu \nu} (A X + 2BX -U) .
 \end{equation}
Our choice of the background metric is a spatially flat FLRW metric, also the scalar field $\phi$ to be spatially homogeneous, which depends only on time $\phi(t)$. Thus we can get the energy density and pressure from the energy-momentum tensor as 
  \begin{equation}
   \rho=T_{0 0}=AX+3BX^2+U  
 \end{equation}
  \begin{equation}
p=T_{i i}=AX+BX^2-U.
\end{equation}
Einsteins' equations of motion are 
\begin{equation}\label{2.15}
   3H^2=\rho  
 \end{equation}
  \begin{equation}\label{2.16}
   \dot\rho+3H(\rho+p)=0  
\end{equation}
where dot represents the derivative with respect to the cosmic time and H is the Hubble parameter. Using eq.(\ref{2.15}) and eq.(\ref{2.16}), we get 
 \begin{equation}
2\dot{H}+3H^2=-p.
 \end{equation}
This leads us to the scalar field equation 
 \begin{equation}
\Ddot{\phi}(A+6BX)+3H\dot{\phi}(A+2BX)-A'X-3B'X^2-U'=0
 \end{equation}
where prime denotes the derivative with respect to $\phi$.
\section{Constant roll Inflation}\label{sec:cri}
The inflaton field $\phi$ is the  only degree of freedom that we have in this model.  We assume that field  $\phi$ satisfies the constant roll condition as
\begin{equation}
\Ddot{\phi}=\beta H\dot\phi,
\end{equation}
where $\beta$ is the constant roll parameter. And this condition is about the same as the slow roll condition where $\ddot \phi \simeq0$ if $\beta\ll1$. We also assume that the field $\phi$ satisfies  $\frac{{\dot \phi}^2}{2}\ll U(\phi)$ during the initial phase of the inflation. For that phase, we are defining slow roll  parameters as  \cite{Odintsov:2019ahz},
\begin{equation}\label{3.2}
\epsilon_{1}=-\frac{\dot H}{H^2},\quad \epsilon_{2}=-\frac{\Ddot \phi}{H\dot \phi},\quad
\epsilon_{3}=\frac{\dot F}{2HF},\quad
\epsilon_{4}=\frac{\dot E}{2HE}
\end{equation}
where 
\begin{equation}
F=\frac{\partial \mathcal{L}}{\partial R}, \quad E=-\frac{F}{2X}\left(X\frac{\partial \mathcal{L}}{\partial X}+2X^2{\frac{\partial^2 \mathcal{L}}{\partial X^2}}\right).
\end{equation}
As we can see, constant roll condition fixes $\epsilon_2 = -\beta$. And we are working in Einstein frame where $F=\frac{1}{2}$, from which we get $\epsilon_3=0$. We may express observable quantities like the scalar spectral index $n_{s}$ and the tensor to scalar ratio r in terms of the slow roll parameters as \cite{Noh:2001ia,Hwang:2002fp,Hwang:2005hb} 
\begin{equation}\label{3.4}
    n_{s}=1-2\frac{(2\epsilon_{1}-\epsilon_{2}-\epsilon_{3}+\epsilon_{4})}{1-\epsilon_{1}}
  \end{equation} 
    \begin{equation}\label{3.5}
    r=4 |\epsilon_{1}| C_{s}  
\end{equation}
where $C_{s}$ is the speed of sound wave of primordial perturbations, expressed by 
\begin{equation}\label{3.6}
    C_{s}^2=\frac{\mathcal{L}_{X}}{\mathcal{L}_{X}+2X\mathcal{L}_{XX}} =\frac{A-B{\dot \phi}^2}{A-3B{\dot \phi}^2},
\end{equation}
which follows, $0<C_s^2<1$. The scalar power spectrum up to first order in slow-roll parameters is expressed as \cite{Lyth:1998xn},
\begin{equation}\label{3.7}
   P_{s} \approx \frac{H^2}{8\pi^2 \epsilon_{1}},
\end{equation}
Inserting constant roll condition into eq.(2.9) and we obtain 
\begin{equation}
  4H\dot{\phi}A(\beta +3)-12H \dot{\phi}^3 B(\beta +1)+A'\dot{\phi}^2-3B'\dot{\phi}^4-4U'=0
\end{equation}
Using eq.(2.4) and eq.(2.6), we obtain an approximate result of H upto second order in terms of $\dot{\phi}^2/2U$ 
as 
\begin{equation}
H=\sqrt{\frac{U}{3}\left(1-A\left(\frac{\dot{\phi}^2}{2U}\right)+3BU\left(\frac{\dot{\phi}^2}{2U}\right)^2\right)}\approx\sqrt{\frac{U}{3}}\left(1-\frac{A}{2}\left(\frac{\dot{\phi}^2}{2U}\right)+\frac{1}{2}\left(3BU-\frac{A^2}{4}\right)\left(\frac{\dot{\phi}^2}{2U}\right)^2\right)
\end{equation}
Next, we substitute this approximated result of H into eq.(3.8), we get
\begin{equation}\label{3.10}
\frac{1}{2}A'\dot{\phi}^2+(\beta +3)A\sqrt\frac{U}{3}\dot{\phi}-U'=0,
\end{equation}
keeping only terms up to $O\left(\dot{\phi}^2\right)$ as higher power terms are negligible. Now the eq.(\ref{3.10}) is a quadratic equation of $\dot{\phi}$, and the solutions of this equation are : 
\begin{equation}
    \dot{\phi} =\begin{cases}
  \frac{-\sqrt{2 A' U'+\frac{1}{3} (\beta +3)^2 A^2 U}-(\beta +3) A \sqrt{\frac{U}{3}}}{A'}, 
  \\
    \frac{\sqrt{2 A' U'+\frac{1}{3} (\beta +3)^2 A^2 U}-(\beta +3) A \sqrt{\frac{U}{3}}}{A'}
    \end{cases}
\end{equation}  
Here, the first solution leads us to an unsolvable set of equations hence we proceed further with the calculation for the second solution of $\dot{\phi}$, which is given by\\
\begin{equation}\label{3.12}
 \dot{\phi}= \frac{\sqrt{2 A' U'+\frac{1}{3} (\beta +3)^2 A^2 U}-(\beta +3) A \sqrt{\frac{U}{3}}}{A'}.
\end{equation}
This equation can be solved by using eq.(\ref{2.11}) with the choice of potential and coupling function.
As we can see, the scalar spectral index, tensor-to-scalar ratio and the scalar power spectrum, are all just the functions of the slow roll parameters as expressed in equations (\ref{3.4}), (\ref{3.5}) and (\ref{3.7}). So first we need to solve the slow roll parameters from the eq.(\ref{3.2}). By using the relations (\ref{2.11}), the general expression (\ref{3.2}) can be simplified as
\begin{equation}\label{5.1}
\epsilon_{1}=\frac{3 X A+6 X^2 B}{X A+3 X^2 B+U}=\frac{6 \gamma  \left(\gamma  {\dot\phi}^4 \epsilon -4 \delta  {\dot\phi}^2\right)}{16 \delta ^2 \phi +\gamma  {\dot\phi}^2 \left(3 \gamma  {\dot\phi}^2 \epsilon -8 \delta \right)}
\end{equation}
\begin{equation}\label{5.2}
\epsilon_{2}=-\beta
\end{equation}
\begin{equation}\label{5.3}
\epsilon_{3}=0
\end{equation}
\begin{align}\label{5.4}
\epsilon_{4} &= \frac{\sqrt{3}}{2} \frac{\dot{\phi}(A'+6B'X)+12\beta BHX}{(A+6BX)\sqrt{AX+3BX^2+U}}  \nonumber \\ 
 &=  \frac{\gamma  \dot{\phi} \left(3 \beta  \dot{\phi} \epsilon  \sqrt{\phi} \sqrt{\epsilon  \phi+1} \sqrt{16 \delta ^2 \phi +\gamma  \dot{\phi}^2 \left(3 \gamma  \dot{\phi}^2 \epsilon -8 \delta \right)}+4 \sqrt{3} \sqrt{\delta } \epsilon  \phi  \left(4 \delta -3 \gamma  \dot{\phi}^2 \epsilon \right)+2 \sqrt{3} \sqrt{\delta } \left(4 \delta -3 \gamma  \dot{\phi}^2 \epsilon \right)\right)}{\sqrt{\phi } \sqrt{\epsilon  \phi +1} \left(3 \gamma  \dot{\phi}^2 \epsilon -4 \delta \right) \sqrt{16 \delta ^2 \phi +\gamma  \dot{\phi}^2 \left(3 \gamma  \dot{\phi}^2 \epsilon -8 \delta \right)}},
\end{align}
similarly, the speed of sound wave of primordial perturbations can be obtained from (\ref{3.6}) as
\begin{equation}
C_s^2=\frac{A-B{\dot \phi}^2}{A-3B{\dot \phi}^2} = \frac{\gamma  {\dot{\phi}}^2 \left(\gamma  {\dot{\phi}}^2 \epsilon -8 \delta \right)-16 \delta ^2 \phi (t)}{\gamma  {\dot{\phi}}^2 \left(\gamma  \left({\dot{\phi}}^2-8\right) \epsilon -8 \delta \right)-16 \delta ^2 \phi (t)}
\end{equation}
and scalar power spectrum from the eq.(\ref{3.7})
\begin{equation}
    P_{s}\approx\frac{H^2}{8\pi^2 \epsilon_{1}}=\frac{1}{72\pi^2}\frac{(AX+3BX^2+U)^2}{AX+2BX^2}=\frac{\left(16 \delta ^2 \phi +\gamma  \dot{\phi} ^2 \left(3 \gamma  \dot{\phi} ^2 \epsilon -8 \delta \right)\right)^2}{2304 \pi ^2 \gamma ^3 \delta  \dot{\phi} ^2 \phi  (\epsilon  \phi +1) \left(\gamma  \dot{\phi} ^2 \epsilon -4 \delta \right)}.
\end{equation}
Here all the slow roll parameters, speed of sound wave and the scalar power spectrum are the function of $\dot{\phi}$, which can be solved by using eq.(\ref{3.12}) for our choice the potential $V(\phi)$ and coupling function $G(\phi)$. Then, this $\dot{\phi}$ can be substituted back to these expressions of $\epsilon$'s, $C_s$ and $P_s$.
\subsection{Results}\label{results}
Our choice of the potential $V(\phi)$ and coupling function $G(\phi)$ is 
\begin{equation}\label{x}
V(\phi)= \delta  \phi^2, \quad G(\phi)=\gamma  \phi,
\end{equation}
even though there are other choices for the potential \cite{Das:2020kff} but we have chosen this for simplicity as it leads to a close set of equations and give viable results. Also, we assume $\epsilon=\frac{8\alpha \delta}{\gamma} \ll 1$ in the effective potential term eq.(\ref{2.11}), $8 \alpha \frac{V(\phi)}{G(\phi)} = 8 \alpha \frac{\delta}{\gamma} \phi$.

By using eq.(\ref{x}) and eq.(\ref{2.11}), we get the solution of eq.(\ref{3.12}) in the power series of $\epsilon$,
\begin{equation}\label{4.1}
\dot{\phi}=-\gamma  \epsilon  \phi^2 \left(\frac{\sqrt{3} (\beta +3) \sqrt{\delta }}{2 \gamma ^2}-\frac{\sqrt{3} \sqrt{\delta } \left(-2 \gamma +\beta ^2 \phi+6 \beta  \phi+9 \phi\right)}{2 (\beta +3) \gamma ^2 \phi}\right).
\end{equation}
Here we ignored the higher power of $\epsilon$ as they are very small since $\epsilon\ll1$. 
As a result of the preceding analysis, we get the solution for $\dot{\phi}[\phi]$, depending on $\phi$ and this can be inserted into all the expressions of the slow roll parameters and the number of e-folds.

The number of e-folds can be calculated from the time when a mode k crosses the horizon  until inflation ends and is written as 
\begin{equation}
N=\int_{\phi_{k}}^{\phi_{f}} \frac{H}{\dot{\phi}} \,d{\phi}={\frac{1}{\sqrt{3}}}{\int_{\phi_{k}}^{\phi_{f}} \frac{\sqrt{X A+3 X^2 B+U }}{\dot{\phi}} \,d{\phi}},
\end{equation}
where $\phi_{k}$ and $ \phi_{f}$ are the values of inflaton field $\phi$ at the horizon crossing and at the end of the inflation, respectively. Since $\phi_{k}\gg \phi_{f}$ during the inflation, we get approximated result of the above equation,
\begin{equation}
N\approx-\frac{\left(\frac{2 (\beta +3)^2}{\gamma  \epsilon }+3\right) \log (\phi_f/\phi_k)}{6 (\beta +3)},
\end{equation}
then $\phi_{k}$ can be obtained as 
\begin{equation}\label{5.9}
    \phi_{k}\approx \phi_{f} \cdot e^{\frac{6 (\beta +3) \gamma  N \epsilon }{2 (\beta +3)^2+3 \gamma  \epsilon }} .
\end{equation}
We can solve the field value $\phi_{f}$ at the end of the inflation by demanding the condition $\epsilon_{1}(\phi_{f})=1$. Up to first order in $\epsilon$, we get 
\begin{equation}
    \phi _f= \frac{1}{\epsilon }
\end{equation} \\
This is value of field $\phi$ at end of the inflation depending on $\epsilon$ and we know $\epsilon={8\alpha\delta}/{\gamma}$. By putting the value of $\phi_f$ and N (range of 50-70) into eq.(\ref{5.9}), the value of $\phi_k$ can be obtained. Putting this value into equations (\ref{5.1}), (\ref{5.2}), (\ref{5.3}) and (\ref{5.4}), the slow roll parameters can be calculated in terms of N, $\beta$ and coupling parameters.
\begin{equation}\label{4.7}
  \epsilon_{1}=-\frac{9 \gamma  \epsilon }{8 (\beta +3)^2} +\frac{27 \gamma ^2 \epsilon ^2 (16 \beta  N+48 N+31)}{128 (\beta +3)^4}
\end{equation}
\begin{equation}\label{4.8}
\epsilon_{2}=-\beta
\end{equation}
\begin{equation}\label{4.9}
\epsilon_{3}=0
\end{equation}
\begin{equation}\label{4.10}
\epsilon_{4}=\frac{9 \epsilon  \left(-\beta  \gamma +2 \sqrt{2} \beta  \gamma +6 \sqrt{2} \gamma \right)}{16 (\beta +3)^2}+\frac{9 \epsilon ^2 \left(87 \beta  \gamma ^2+2 \sqrt{2} (\beta +3) \gamma  (-45 \gamma -28 \beta  \gamma  N-84 \gamma  N)\right)}{256 (\beta +3)^4}.
\end{equation}
Finally, the scalar spectral index and the tensor-to-scalar ratio are obtained  by substituting eq.(\ref{4.7}) - (\ref{4.10}) into eq.(\ref{3.4}) and eq.(\ref{3.5}) 
\begin{equation} \label{eqn:r}
\begin{aligned}
r = 4 \left(1-\frac{3 \gamma ^2 \epsilon ^2}{16 \left((\beta +3)^2 \delta \right)}+\frac{9 \gamma ^3 \epsilon ^3 (8 \beta  N+24 N+17)}{128 (\beta +3)^4 \delta }\right) \Bigg| \frac{27 (16 \beta  N+48 N+31) \gamma ^2 \epsilon ^2}{128 (\beta +3)^4}-\frac{9 \gamma  \epsilon }{8 (\beta +3)^2}\Bigg|
 \end{aligned}
\end{equation}
\begin{equation}\label{eqn:ns}
\begin{aligned}
 n_{s} = & 1-2 \beta-\frac{9 (3 \beta +4) \gamma  \epsilon }{8 (\beta +3)^2}-\frac{9 \gamma  \epsilon }{2 \sqrt{2} (\beta +3)}+\frac{63 \gamma ^2 \epsilon ^2 (4 (\beta +3) N+9)}{32 \sqrt{2} (\beta +3)^3}\\&-\frac{27 \gamma ^2 \epsilon ^2 \left(109 \beta +32 \left(\beta ^2+5 \beta +6\right) N+148\right)}{128 (\beta +3)^4}
 \end{aligned}
 \end{equation}
and speed of sound is given by 
\begin{equation}
    C_s=\left(1-\frac{3 \gamma ^2 \epsilon ^2}{16 \left((\beta +3)^2 \delta \right)}+\frac{9 \gamma ^3 \epsilon ^3 (8 \beta  N+24 N+17)}{128 (\beta +3)^4 \delta }\right)
\end{equation}
which shall follow the condition,
\begin{equation}
0 < C_s <1 \label{6.4}.
\end{equation}
Now, to match our results with the observations, we choose $\epsilon=10^{-7}$ with $0.006<\beta<0.018$ and the number of e-folds($N$) from 50 to 70 satisfying $\epsilon=8\frac{\alpha \delta}{\gamma}\ll1$ condition. For this parameter regime, eq.(\ref{6.4}) implies $\delta$ must be much greater than $10^{-5}$ and hence we choose $\delta=1$. With the fixed values of $\epsilon$ and $\delta$, the only remaining parameters are $N$, $\beta$ and $\gamma$. Then in Fig.$1$, we fix $\gamma=2.5\times10^{5}$  while varying $\beta$ and $N$ within the allowed range and plot $r$ vs $n_s$ results using eq.(\ref{eqn:ns}) and eq.(\ref{eqn:r}). Here, we also notice that as we vary $N$ from $50$ to $70$, the value of $r$ increases while increasing $\beta$ lead to a smaller value of $n_s$. Then in Fig.$2$, we overlap our analytical results over Planck 2018 data and show that the results are well in agreement with the observation. Following the same procedure as Fig.$1$, we make Fig.$3$, Fig.$5$, Fig.$7$ and Fig.$9$ for different values of $\gamma$ as shown. Accordingly, all these figures are then overlapped just like Fig.$2$ over the Planck data as shown in Fig.$4$, Fig.$6$, Fig.$8$ and Fig.$10$. With this comparison with the observation, we find that an increased value of $\gamma$ lead to mismatching with the observations thus constraining this parameter as well.

\begin{figure}[hbt!]
  \centering
  \begin{minipage}[b]{0.4\textwidth}
    \includegraphics[width=\textwidth]{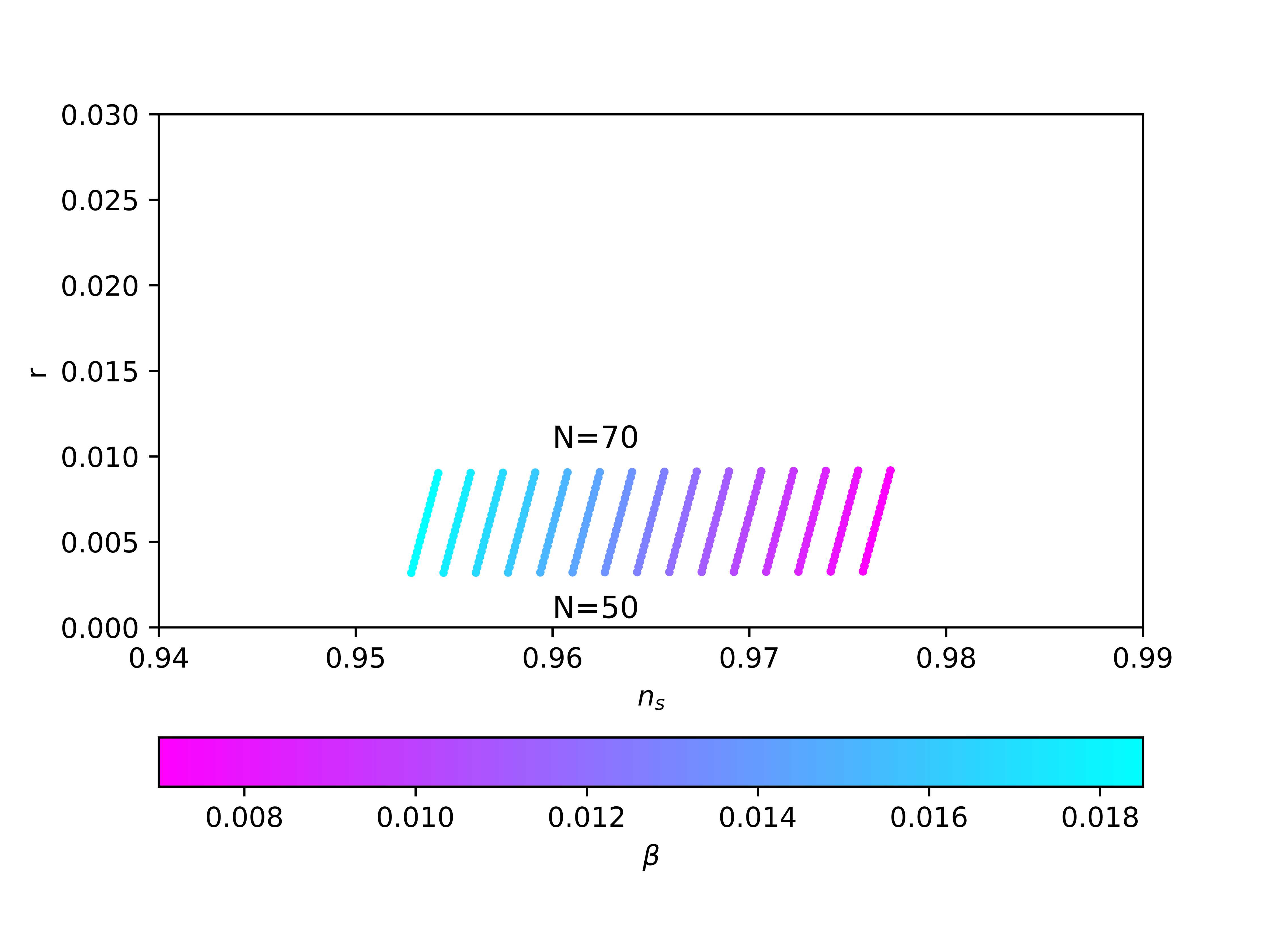}
    \caption{$\gamma = 2.5\times10^5$}
 \end{minipage}
 \hfill
\begin{minipage}[b]{0.4\textwidth}
    \includegraphics[width=\textwidth]{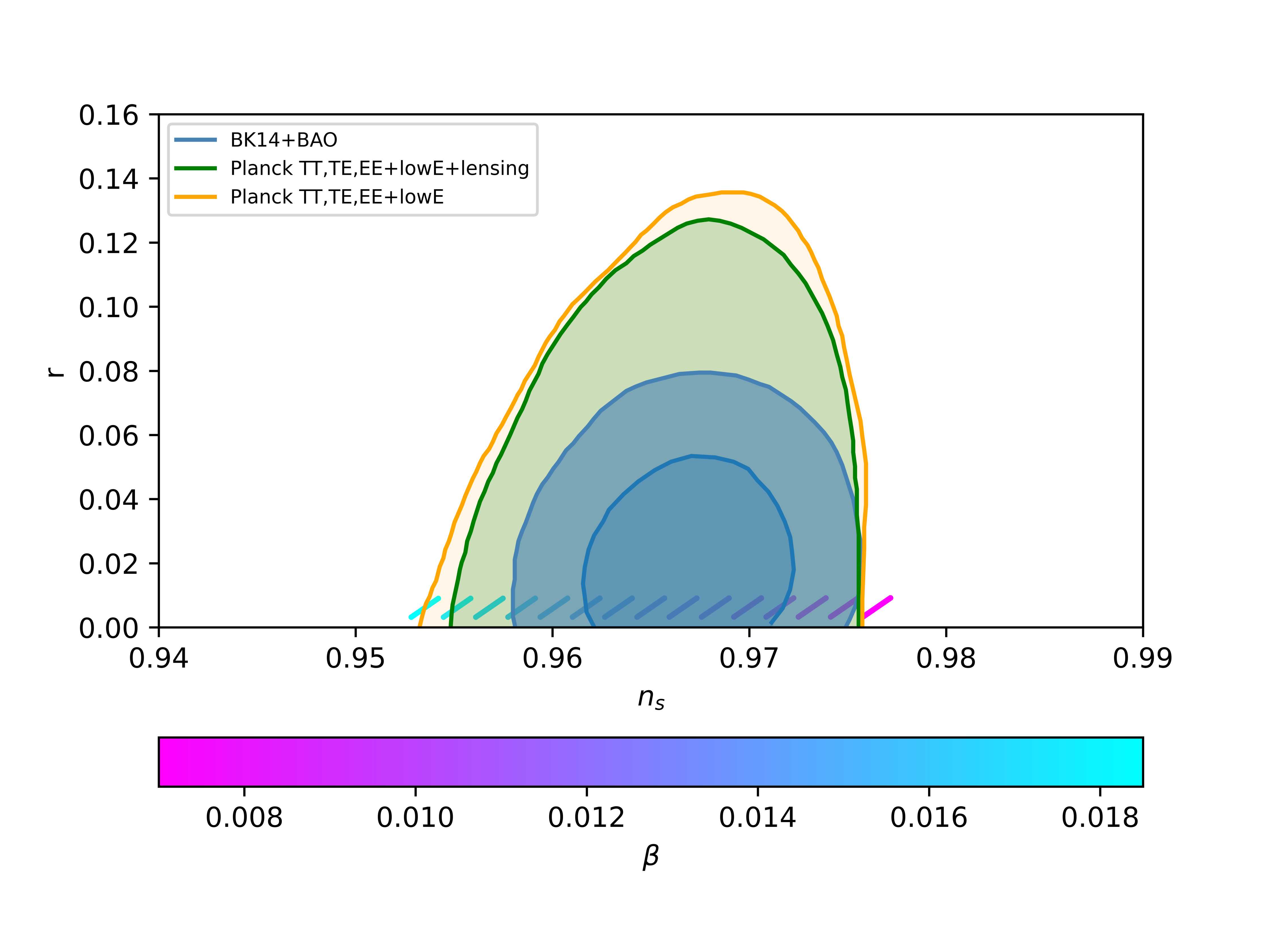}
    \caption{$\gamma = 2.5\times10^5$}
  \end{minipage}
\end{figure}
\begin{figure}[hbt!]
  \centering
  \begin{minipage}[b]{0.4\textwidth}
    \includegraphics[width=\textwidth]{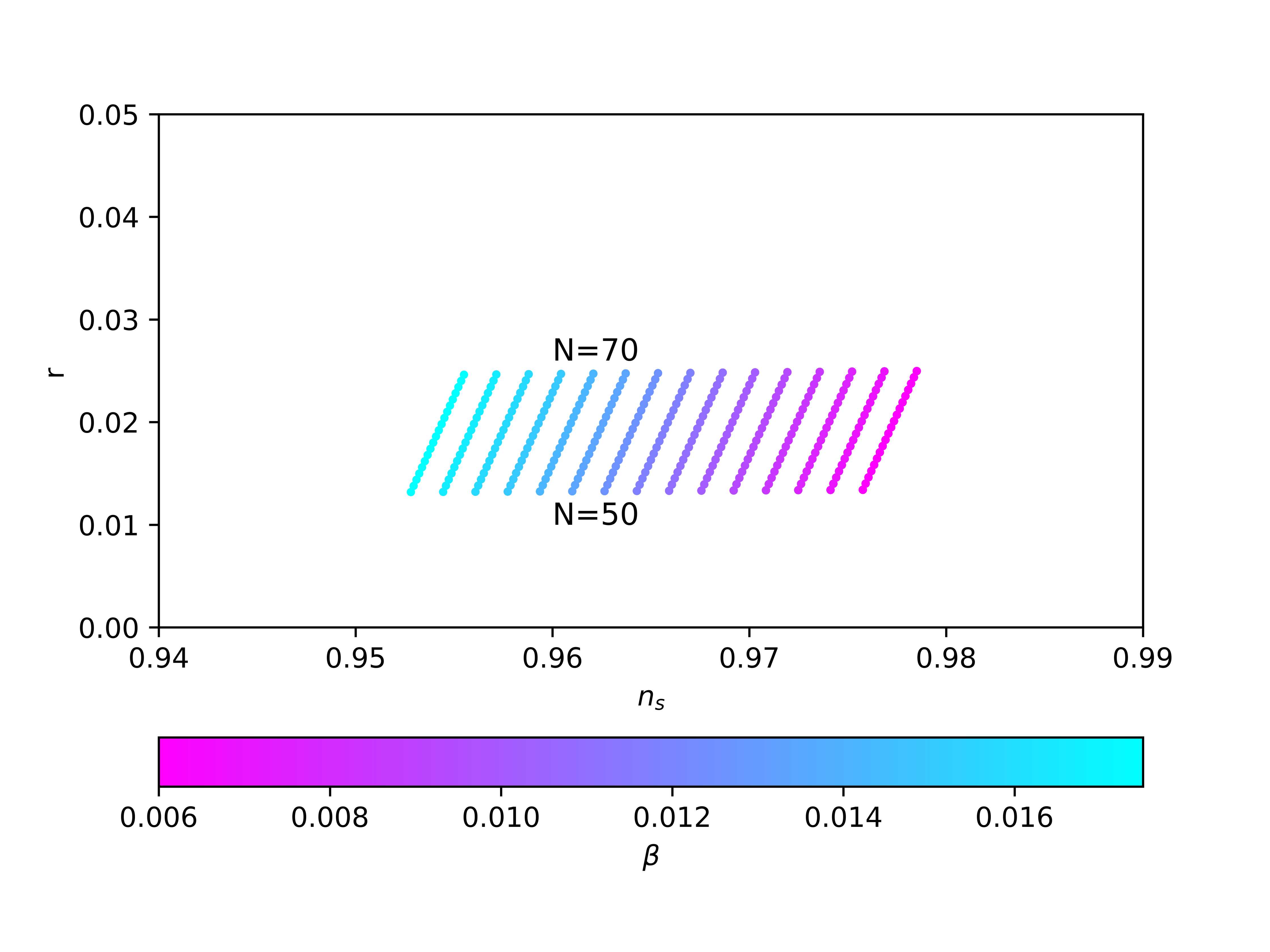}
    \caption{$\gamma = 3.5\times10^5$}
  \end{minipage}
  \hfill
  \begin{minipage}[b]{0.4\textwidth}
    \includegraphics[width=\textwidth]{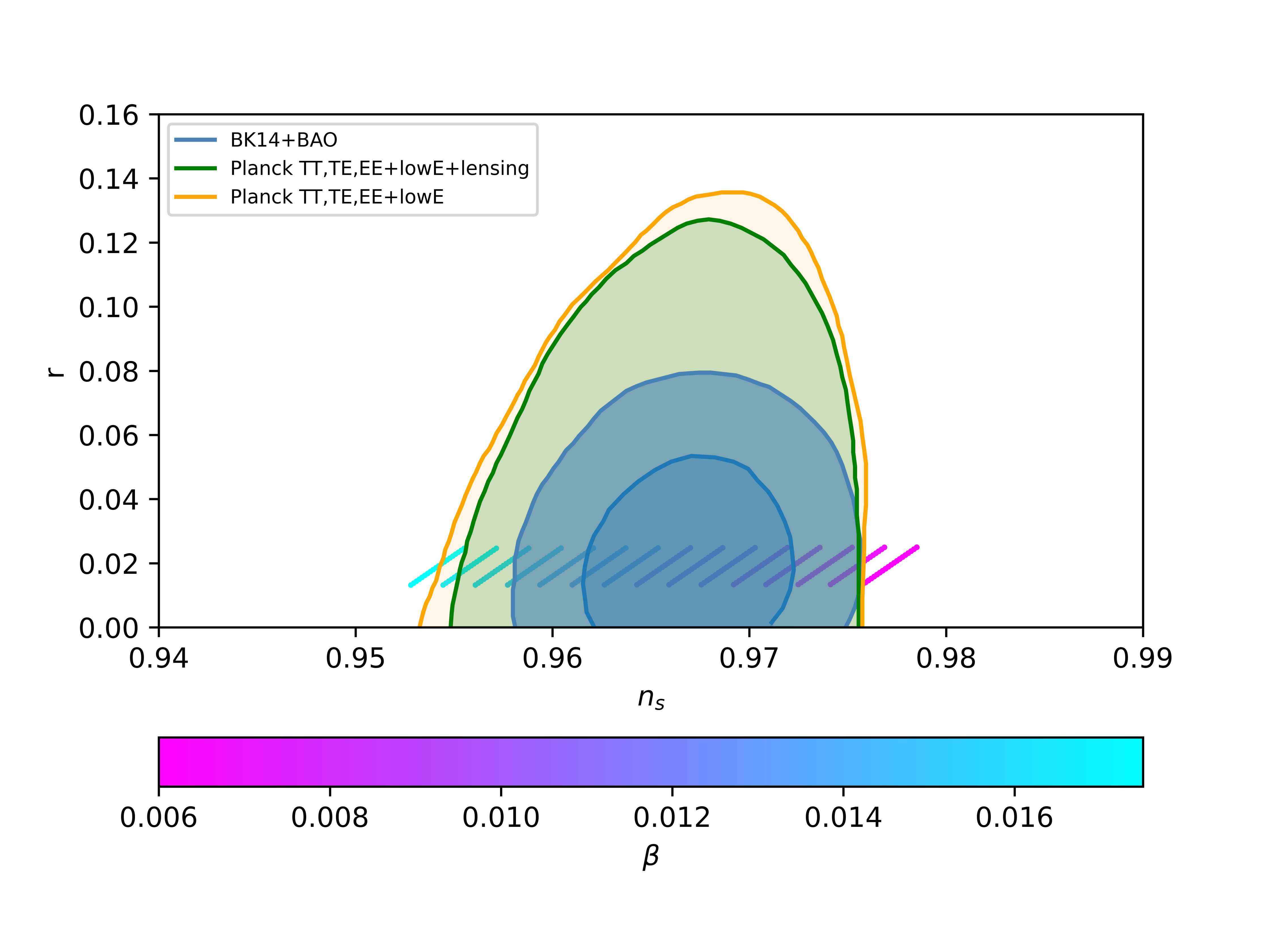}
    \caption{$\gamma = 3.5\times10^5$}
  \end{minipage}
\end{figure}
\begin{figure}[hbt!]
  \centering
  \begin{minipage}[b]{0.4\textwidth}
    \includegraphics[width=\textwidth]{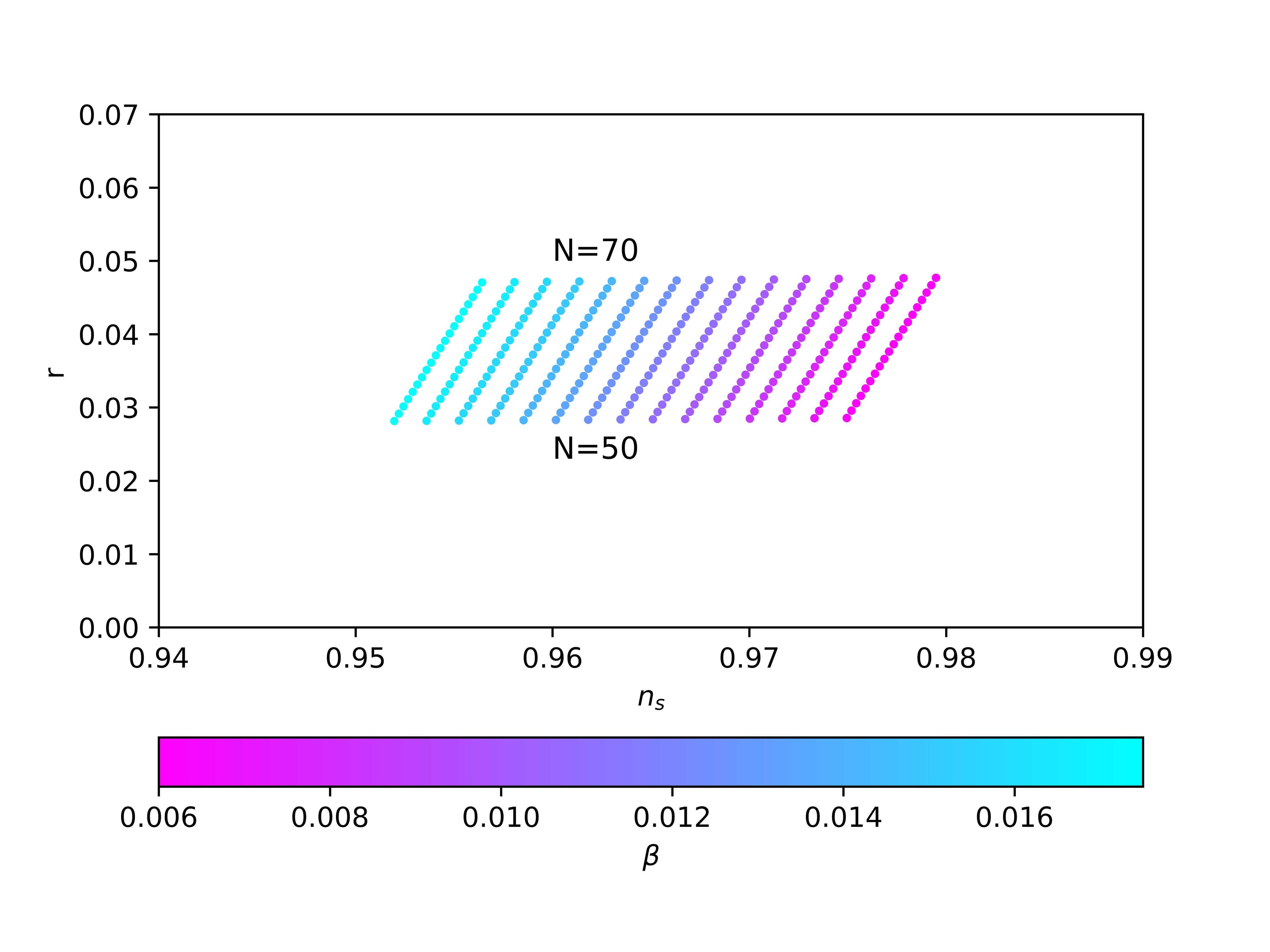}
    \caption{$\gamma = 4.5\times10^5$}
  \end{minipage}
  \hfill
  \begin{minipage}[b]{0.4\textwidth}
    \includegraphics[width=\textwidth]{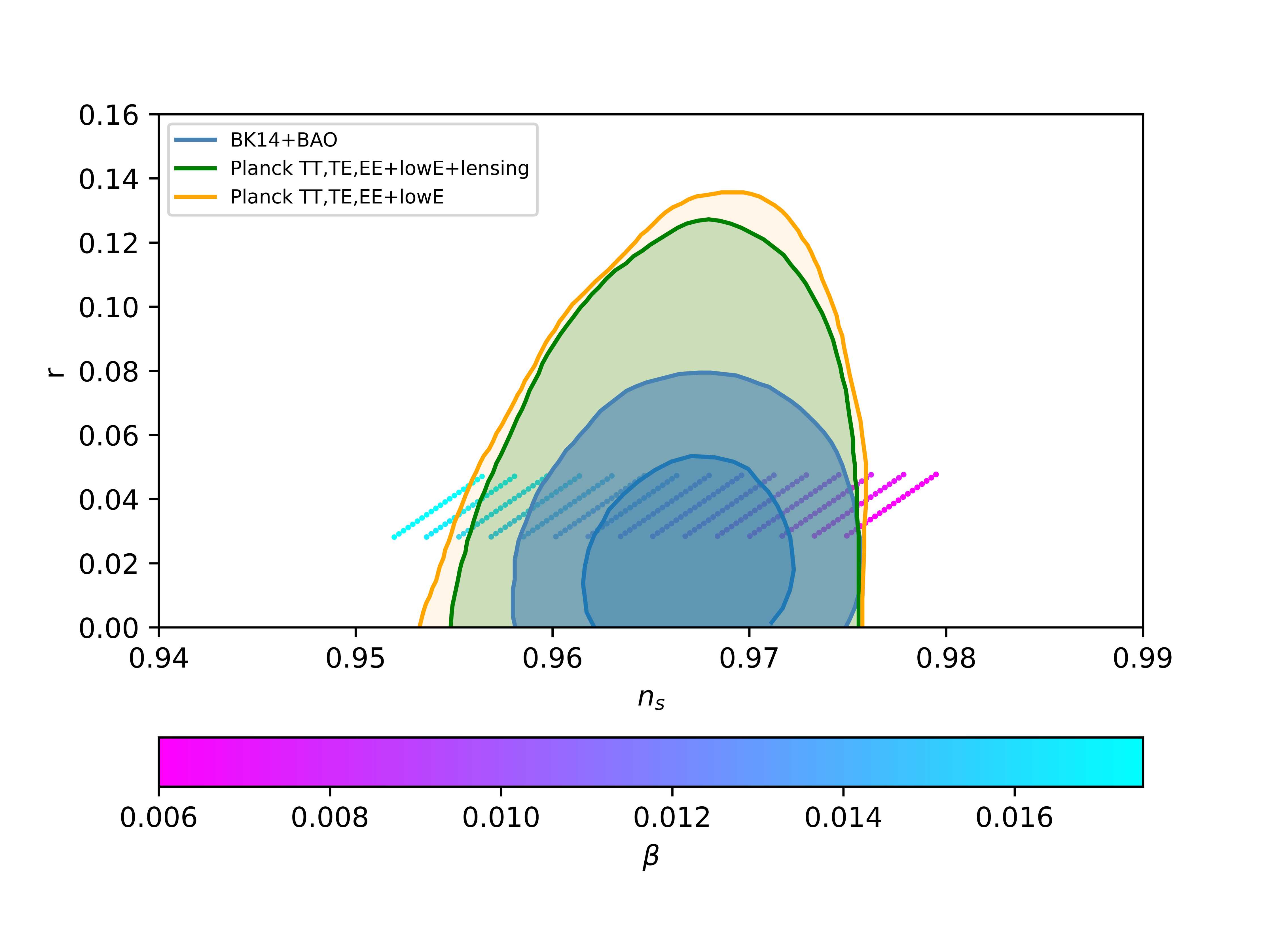}
    \caption{$\gamma = 4.5\times10^5$}
  \end{minipage}
\end{figure}\begin{figure}[hbt!]
  \centering
  \begin{minipage}[b]{0.4\textwidth}
    \includegraphics[width=\textwidth]{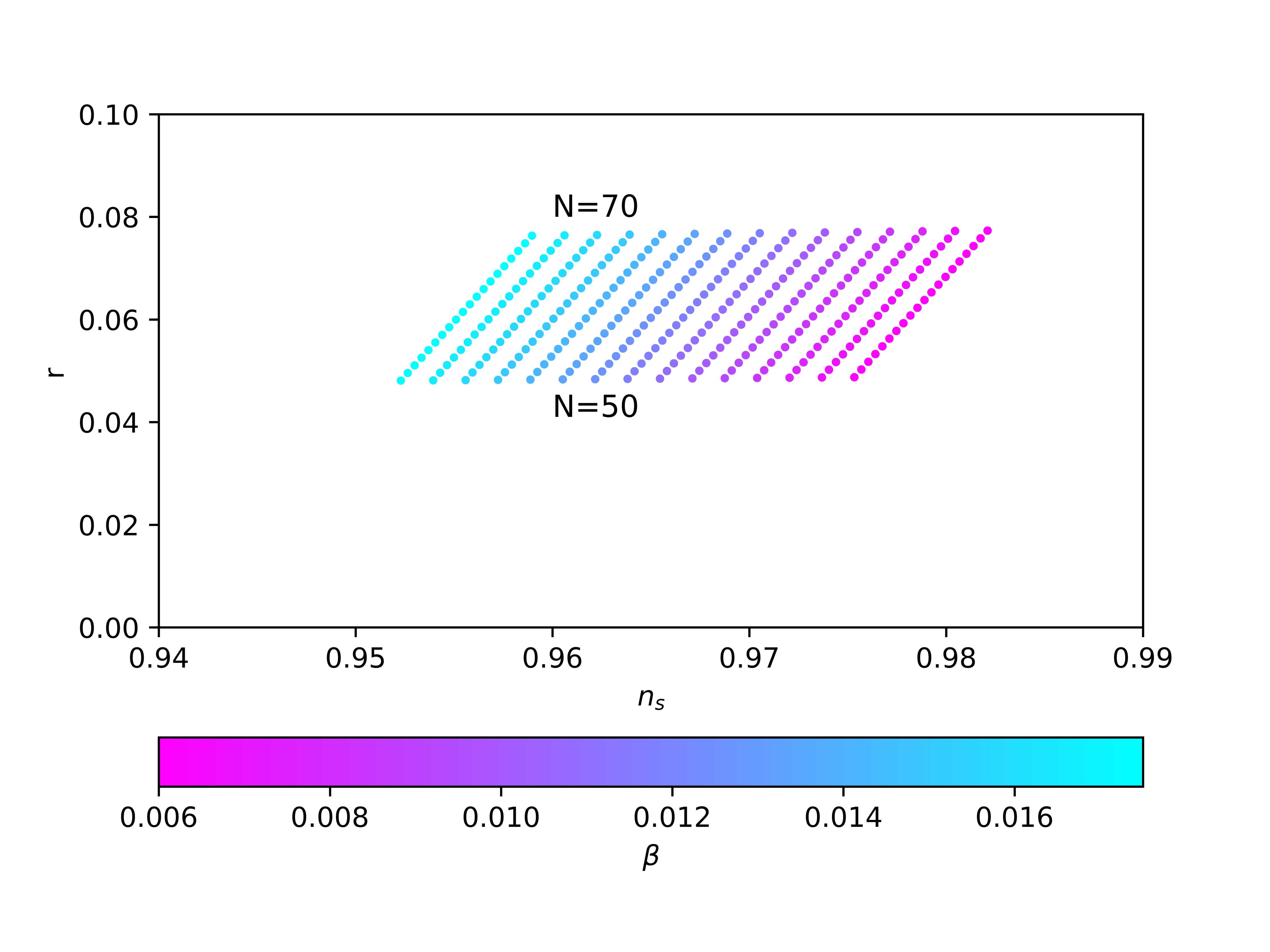}
    \caption{$\gamma = 5.5\times10^5$}
  \end{minipage}
  \hfill
  \begin{minipage}[b]{0.4\textwidth}
    \includegraphics[width=\textwidth]{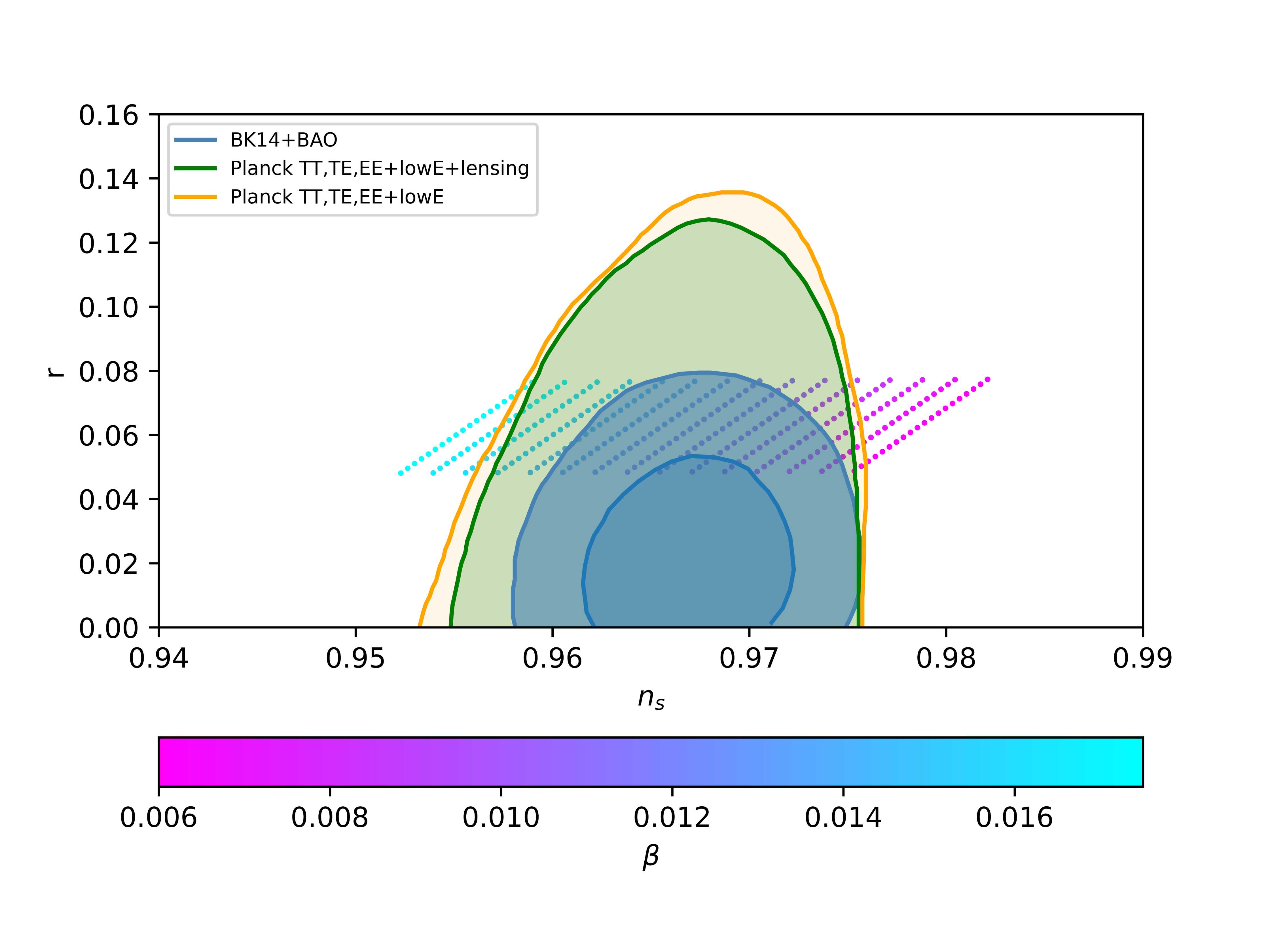}
    \caption{$\gamma = 5.5\times10^5$}
  \end{minipage}
\end{figure}
\begin{figure}[hbt!]
  \centering
  \begin{minipage}[b]{0.4\textwidth}
    \includegraphics[width=\textwidth]{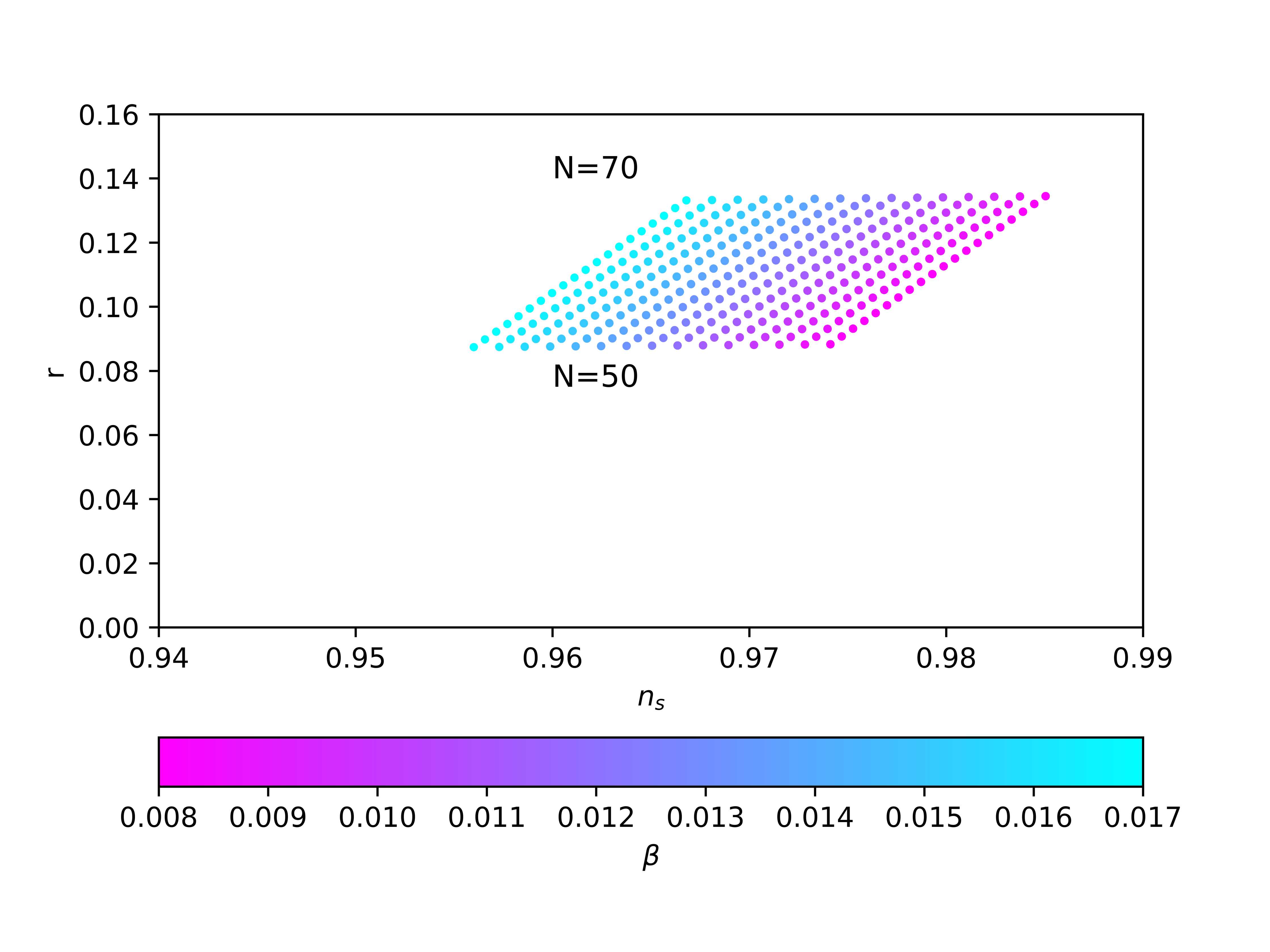}
    \caption{$\gamma = 7\times10^5$}
  \end{minipage}
  \hfill
  \begin{minipage}[b]{0.4\textwidth}
    \includegraphics[width=\textwidth]{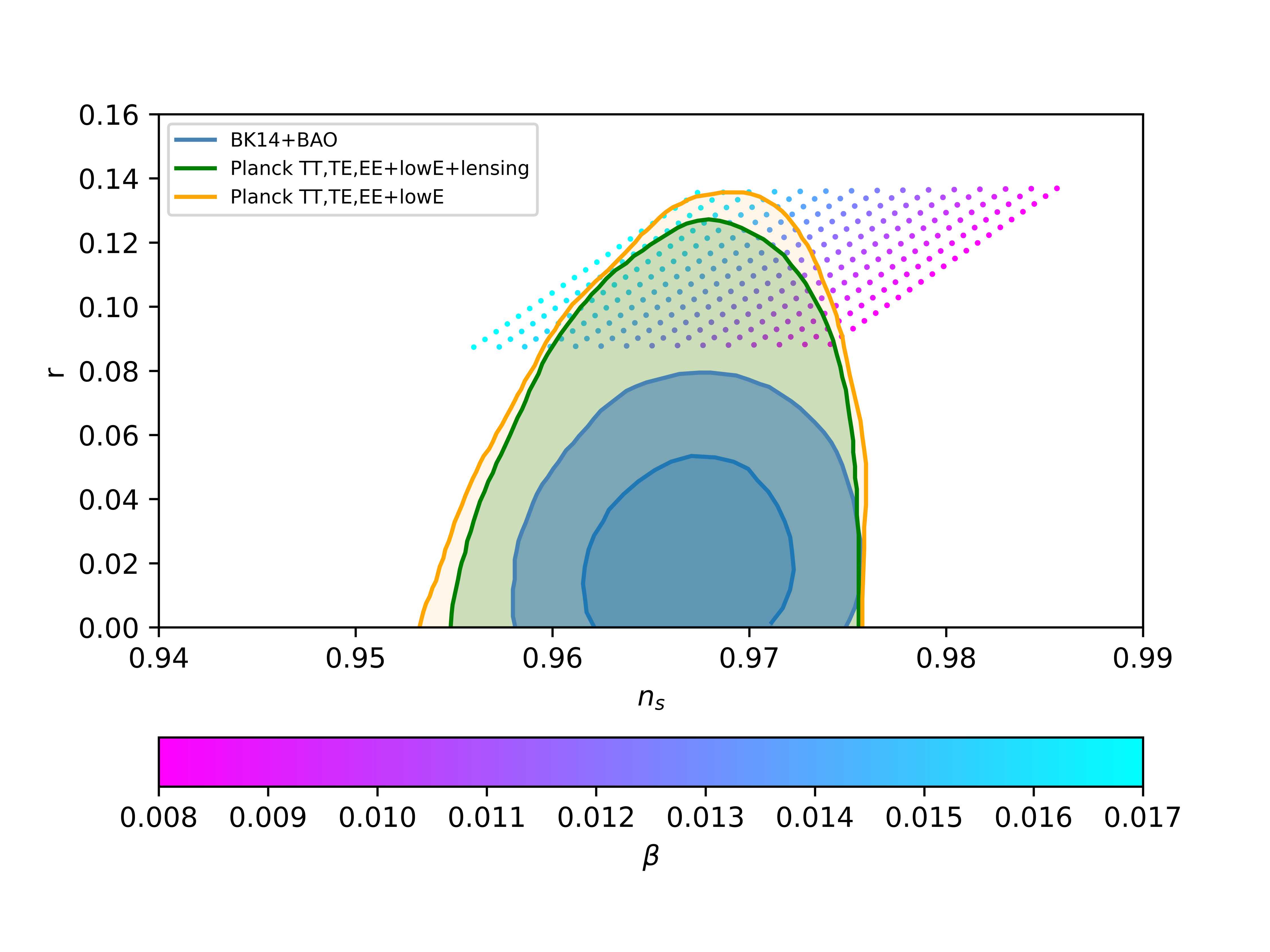}
    \caption{$\gamma = 7\times10^5$}
  \end{minipage}
\end{figure}
\par
\clearpage
\section{Conclusion}\label{sec:conclusion}
\par
In this paper, we have investigated the possibility of constant-roll in the modified gravity theory using Palatini formalism with action containing non-minimal coupling of $R$ and $R^2$ with a scalar field. This formalism helps us in translating $R^2$ coupling term into a higher-order kinetic term with a new potential making it an equivalent scalar-tensor theory in the Einstein frame. Then we proceeded to find the equation of motion for the scalar field  within this frame and using the constant roll condition, we were able to solve the field equation and find the slow roll parameters. Then we defined our potential for the theory and worked out the expressions for tensor-to-scalar ratio and spectral index. Within the parameter regime of our theory, our results are in complete agreement with the Planck 2018 data and the results with the inclusion of comparison with Planck data are shown in the Figs.($1-10$), thus making constant roll a viable scenario for further research in this model. Most agreed values of $n_s$ and $r$ are obtained in the first case. In future, we plan to study the possibility of reheating \cite{Oikonomou:2017bjx} and the production of primordial black hole \cite{Motohashi:2019rhu}  within the Palatini formalism for this model.


\section{Acknowledgement}
This work is partially supported by DST
(Govt. of India) Grant No. SERB/PHY/2021057 and the author RT would like to thank  Abhijith Ajith for his help in making plots.
\begin{center}
 \rule{4in}{0.5pt}\\
 \vspace{-11.5pt}\rule{3in}{0.5pt}\\
 \vspace{-11.5pt}\rule{2in}{0.5pt}\\
\end{center}
\printbibliography[heading=none]
\end{document}